\documentstyle[12pt]{article}

\newcommand{\sbm}[1]{\mbox{\boldmath{$#1$}}}
\begin{document}
\begin{titlepage}
\title{Two-body Pion Absorption on $^3He$ at Threshold}

\author{L. L. Kiang \\
Department of Physics, National Tsing-Hua University \\
Hsinchu, Taiwan, Republic of China
	   \\
	   \\
T.-S. H. Lee \\
Physics Division, Argonne National Laboratory \\
Argonne, Illinois 60439, USA
            \\
            \\
D. O. Riska \\
Department of Physics, University of Helsinki \\
Helsinki 00014, Finland }
\date{}
\maketitle
\vspace{10. mm}
\centerline{Abstract}

It is shown that a satisfactory explanation of the ratio of
the rates of the reactions  $^3He(\pi^-,nn)$ and $^3He(\pi^-,np)$ for
stopped pions is obtained once the effect of the short range two-nucleon
components  of the axial charge operator for the nuclear system is taken
into account.
By employing realistic models for the
nucleon-nucleon interaction in the construction of these
components of the axial charge operator, the predicted ratios
agree with the empirical value to within 10-20\%.

\end{titlepage}
\baselineskip=20pt

\centerline{\bf 1. Introduction}

A natural ansatz for the description of nuclear pion absorption and
production reactions in the near threshold regime, where the reactions
mainly involve $S$-wave pions, is to express the reaction amplitudes
in terms of the axial charge density operator of the nuclear system
[1]. This operator is formed of a single nucleon term and a two-nucleon
"exchange current" term. The latter separates into a long range pion
exchange term, first derived in [2], and a short range term, the most
important part of which can be determined directly from the
nucleon-nucleon interaction [3]. Direct evidence for the short range
component was recently found by the quantitatively successful explanation
of the reaction $pp\rightarrow pp\pi^0$ near threshold [4], to which
the main pion exchange term does not contribute [1]. This important
role of the short range part of the axial exchange current operator
has been substantiated in subsequent work [5,6].

	In this paper we apply the ansatz of describing nuclear $S$-wave
pion
absorption by the axial charge operator to the case of absorption of
stopped negative pions on $^3He$. This reaction, for which the isospin
structure of the nuclear states involved is richer than that in the
case of the reaction $pp\rightarrow pp\pi^0$, is less selective and
therefore the rate also has an important long range
pion exchange contribution. Yet if only this rescattering contribution
is taken into account in addition to the single nucleon operator
an overprediction by more than an order of magnitude of the empirical ratio
[7] of the rates
for production of nn and np pairs results.
The point of this paper is to show that
this problem is resolved by the effect of the
short range contributions to the axial charge operator,
which were recently found in Ref.[1] to provide the
explanation for the cross section for the reaction $pp\rightarrow
pp\pi^0$ near threshold. We show that
when realistic models for the nucleon-nucleon
interaction are employed in the construction of the short range
components of the axial exchange charge operator, a very satisfactory
account of the
recent data [7] on
the predicted ratio of the
two-nucleon absorption processes $^3He(\pi^-,nn)p$ and $^3He(\pi^-,np)n$
is obtained. The short range corrections reduce the predicted value
by the required large factor so that the final predicted ratio
exceeds the empirical value $6.3\pm1.1$ [7] by only 10-20\%, the range of
variation arising from the differences between the potential models
considered.

In section 2 we review the derivation of the $S$-wave
pion absorption operator, as it appears when constructed from the
axial charge operator. The formalism for calculating the absorption
rate for stopped pions by a pair of nucleons in a given eigenchannel
is then presented in section 3. The numerical results and discussions
are given in section 4.

\vspace{0.4 cm}

\centerline{\bf 2. The $S$-wave pion absorption operator}

	We shall take the effective  Lagrangian for the interaction between
low energy pions and nuclei to have the form

\begin{equation} {\cal L}(x) =-\frac{1}{f_\pi} \sbm{A}_\mu(x)\cdot\partial^\mu
\sbm{\phi}(x),
\end{equation}
where $\sbm{\phi}$ is the pion field, $f_\pi$ is the pion decay constant
and $\sbm{A}_\mu$ the axial current
density of the nuclear system.
This Lagrangian density is the direct generalization of Weinberg's
effective Lagrangian for the pion nucleon system [8]. In the
case of a single nucleon the axial current operator has the form

\begin{equation} \sbm{A}_\mu(0)=ig_A\bar u(p')\gamma_\mu\gamma_5
\frac{\sbm{\tau}}{2}u(p),
\end{equation}
and thus in this case Eq.(1) is equivalent to the usual pseudovector
$\pi NN$ coupling model, since by the Goldberger-Treiman relation
the coefficient $g_A/2f_\pi$ can be expressed as

\begin{equation} \frac{g_A}{2f_\pi}=\frac{f_{\pi NN}}{m_\pi},
\end{equation}
where $f_{\pi NN}$ is the pseudovector coupling constant.

	The interaction Eq.(1) implies that $S$-wave pion production
and absorption at threshold is determined by the axial charge operator, which
in
the case of a single nucleon is

\begin{equation} \sbm{A}_0(0)=-g_A\sbm{\sigma}\cdot\sbm{ v}
\frac{\sbm{\tau}}{2},
\end{equation}
where $\sbm{ v}$ is the nucleon velocity operator.

	The most important two-nucleon contribution to the axial
charge operator is found in Ref.[3] to be the $\pi\rho$-exchange operator,
which has the form

\begin{equation} \sbm{ A}_0^{(\pi\rho)}(0)=i\frac{g_A m_\rho^2}{2f_\pi^2}
\frac{\sbm{\sigma}^2\cdot\sbm{ k}_2}{(k_1^2+m_\rho^2)(k_2^2+m_\pi^2)}
(\sbm{\tau}^1\times\sbm{\tau}^2) + (1\leftrightarrow 2).
\end{equation}
Here the symbol $(1\leftrightarrow 2)$ represents interchange of all
the coordinates of the nucleon pair. We denote the fractions of the
momentum $\sbm{ k}$ of the absorbed pion that are imparted to the
two nucleons as $\sbm{ k}_1$ and $\sbm{ k}_2$ ($\sbm{ k}=\sbm{ k}_1
+\sbm{ k}_2$). In the limit $m_\rho\rightarrow $ infinite the $\pi\rho$
exchange
operator Eq.(5) reduces to the pion
exchange operator of Ref.[2]. As noted in Ref.[9], it
in this limit also corresponds to the isospin asymmetric term
($\sbm{\tau}^1\times\sbm{\tau}^2$) in the
usual form for the $S$-wave pion rescattering operator calculated
using
the usual pseudovector $\pi NN$ coupling and the conventional effective
Hamiltonian for $S$-wave $\pi N$ scattering [10,11]:

\begin{equation} T=-\frac{8\pi i}{\sqrt2\omega_\pi}\frac{f_{\pi NN}}{m_\pi}
\frac{\sbm{\sigma}^2\cdot\sbm{ k}_2}{m_\pi^2+k_2^2}\\
\lbrace\frac{\lambda_1}{m_\pi}\sbm{\tau}^2-i\frac{\lambda_2}{2m_\pi^2}
(\omega_\pi+\omega_2)\sbm{\tau}^1\times\sbm{\tau}^2\rbrace
+(1\leftrightarrow 2).\end{equation}
Here $\omega_2$ is the energy of the exchanged pion
and $\omega_\pi$ that of the initial pion.
The coefficients $\lambda_{1,2}$ are the following
combinations of the two
$S$-wave pion-nucleon scattering lengths $a_{1,3}$:

\begin{equation}
\lambda_1=-\frac{1}{6}m_\pi(a_1+2a_3),\quad\lambda_2=\frac{1}{6}m_\pi(a_1-a_3).
\end{equation}
The $\pi\rho$ component Eq.(5) of the axial exchange charge operator
is obtained by choosing the following values for the coupling
constants $\lambda_{1,2}$ [9]:

\begin{equation} \lambda_1=0,\quad\lambda_2=\frac{m_\pi^2}{16\pi f_\pi^2}
\simeq 0.04, \end{equation}
which agree well with the values extracted from $\pi N$ phase shift
analyses [12,13].

	The short range components of the axial exchange charge
operator that were found to be important for the cross section
of the reaction $pp\rightarrow pp\pi^0$ near threshold correspond
to two-nucleon operators that arise when the axial field excites
intermediate negative energy components of the nucleon (nucleon-
antinucleon "pairs"), which are then de-excited by the short range components
of the nucleon-nucleon interaction (Fig.1) [1].
Although these contributions
are most readily derived by employment of explicit boson exchange models,
the corresponding operators can be constructed directly from any existing
nucleon-nucleon interaction model, if that is expressed in terms
of Lorentz invariant spin-amplitudes as e.g. the Fermi invariants.
This construction has been presented in detail in Refs.[3,14].

In the decomposition of the nucleon-nucleon interaction in terms of
Fermi invariants the short range components are contained in the
amplitudes $SVTA$ (the $P$-amplitude contains the long range pion
exchange term, which has to be dropped once the $\pi\rho$ exchange
contribution above is included explicitly in the pion rescattering
contribution). Numerically the most important of these short range axial
exchange charge operators are those that are associated with the
scalar ($S$) and vector ($V$) invariants:

\begin{eqnarray}
&& \sbm{ A}^0(S)=\frac{g_A}{m_N^2}\lbrace[v_S^+(\sbm{ k}_2)
\sbm{\tau}^1+v_S^-(\sbm{ k}_2)\sbm{\tau}^2]
\sbm{\sigma}^1\cdot\sbm{ P}_1\cr\cr&&
+\frac{i}{2}v_S^-(\sbm{ k}_2)\sbm{\tau}^1
\times\sbm{\tau}^2 \sbm{\sigma}^1
\cdot\sbm{ k}_2\rbrace +(1\leftrightarrow 2),\cr\cr. &&
\sbm{ A}^0(V)=\frac{g_A}{m_N^2}\lbrace[v_V^+(\sbm{ k}_2)
\sbm{\tau}^1+v_V^-(\sbm{ k}_2)\sbm{\tau}^2][\sbm{\sigma}^1\cdot\sbm{ P}_2
+\frac{i}{2}\sbm{\sigma}^1\times\sbm{\sigma}^2\cdot\sbm{ k}_2]\cr\cr &&
+\frac{i}{2}v_V^-(\sbm{ k}_2)\sbm{\tau}^1\times\sbm{\tau}^2\sbm{\sigma}^1
\cdot\sbm{ k}_2\rbrace+(1\leftrightarrow 2),\cr\cr &&
\end{eqnarray}
Here we have defined the nucleon momentum operators $\sbm{ P}_{1,2}$
for the two nucleons as $(\sbm{ p}_{1,2}+\sbm{ p}_{1,2}^{'})/2$ respectively.

The potential components $v_S$ and $v_V$ can be given a direct
interpretation as scalar and vector meson exchange potentials. In
the case of a single meson exchange model these would have the
expressions

\begin{equation} v_S^\pm(\sbm{ q})=-\frac{(g_S^\pm)^2}{(m_S^\pm)^2+q^2},
\end{equation}
\begin{equation} v_V^\pm(\sbm{ q})=\frac{(g_V^\pm)^2}{(m_V^\pm)^2+q^2}.
\end{equation}

Here $m_S^\pm$ and $m_V^\pm$ are the masses of the isospin $0$ and $1$
scalar and vector mesons respectively, and $g_S^\pm$ and $g_V^\pm$ are
the corresponding scalar- and vector meson-nucleon coupling constants.
Below we shall use the notation $\sigma$ and $\delta$ to denote the
isospin $0$ and $1$ scalar mesons respectively.

Using the expressions above in defining
the effective Lagrangian Eq.(1), the transition T-matrix
for the absorption of a pion with charge $\alpha$ and momentum $\sbm{k}$
by a nuclear system can now be written as
\begin{eqnarray}
S_{fi} = \delta_{fi} - i 2\pi\delta ^{(4)}(P_f - P_i - k)
< \Psi_f \mid A_{\alpha} (\sbm{k}) \mid \Psi_i >
\end{eqnarray}
where the axial charge density is formed as the sum of one- and two-
nucleon contributions
\begin{eqnarray}
A_{\alpha}(\sbm{k}) = A^{(1)}_{\alpha}(\sbm{k}) + A^{(2)}_{\alpha}(\sbm{k}).
\end{eqnarray}
We shall here restrict the expressions to the limit $\sbm{k}\rightarrow 0$.
The choice of the kinematic variables for a two-nucleon system are indicated in
Fig.1.
For the one-body operator $A^{(1)}_{\alpha}(\sbm{k})$
we obtain from Eq.(4)
the following matrix element
in the momentum representation
\begin{eqnarray}
<\sbm{p}^\prime_1 \mid A^{(1)}_{\alpha} \mid \sbm{p}_1 >&=&C (2\pi)^3
(-1)^\alpha \tau_\alpha^1 \nonumber \\
&\times&\frac{\sbm{\sigma}^1\cdot
\sbm{P}_1 }{m_N}.
\end{eqnarray}
Here the coefficient $C$ has been defined as
\begin{equation}
C=i\frac{\omega_{\pi}(k)}{(2\pi)^{9/2}}
\frac{1}{\sqrt{2\omega_{\pi}(k)}}\frac{f_{\pi NN}}
{m_{\pi}}.
\end{equation}

In terms of the kinematical variables in Fig.1, the contribution of
pion $S$-wave rescattering term Eq.(6) to two-body
axial exchange charge operator $A^{(2)}_{\alpha}(\sbm{k})$
is given by the following matrix element
\begin{eqnarray}
<\sbm{p}^\prime_1 \sbm{p}^\prime_2 \mid A^{(\pi)}_{\alpha}(\sbm{k})
\mid \sbm{p}_1 \sbm{p}_2 >= C\frac{8\pi}{\omega_{\pi}(k)}
 v_{\pi}(\sbm{q})(-1)^{\alpha} \nonumber \\
\times\left\{\frac{\lambda_1}{m_\pi}\tau_\alpha^2-
\frac{(\omega_{\pi}(q)+\omega_{\pi}(k))
\lambda_2}{2m_{\pi}^2}i[\sbm{\tau}^1\times
\sbm{\tau}^2]_\alpha \right\}
 \sbm{\sigma}^2\cdot \sbm{q} + ( 1 \leftrightarrow 2 )
{}.
\end{eqnarray}
Here $v_\pi$ represents the pion exchange interaction
without coupling constants:
\begin{equation}
v_{\pi}(\sbm{q}) = \frac{1}{\sbm{q}^2+m_\pi^2}F^2_{\pi}(\sbm{q}), \nonumber
\end{equation}
where $F_\pi(\sbm{q})$ represents the vertex form factor, which here has
been taken to be the same at the $S$- and the $P$-wave vertices.

{}From the short range components of the axial charge
operators defined in  Eqs. (9) we obtian the following explicit contributions
for the scalar and vector
exchanges mechanisms
($\sigma,\omega$ for isoscalar and $\delta,\rho$ for isovector)
:
\begin{eqnarray}
< \sbm{p}^\prime_1 \sbm{p}^\prime_2 \mid A^{(\sigma)}_{\alpha}(\sbm{k})
\mid \sbm{p}_1 \sbm{p}_2 >& =& C(-1)^\alpha\tau_{\alpha}^1
 \frac{-1}{m_N^2}v_{\sigma}(\sbm{q})
\sbm{\sigma}^1\cdot \sbm{P}_1 +( 1 \leftrightarrow 2 ),
 \\
< \sbm{p}^\prime_1 \sbm{p}^\prime_2 \mid A^{(\delta)}_{\alpha}(\sbm{k})
\mid \sbm{p}_1 \sbm{p}_2 >& =& C(-1)^\alpha\tau_{\alpha}^2
 \frac{-1}{m_N^2}v_{\delta}(\sbm{q}) \sbm{\sigma}^1\cdot
\sbm{P}_1 +(1 \leftrightarrow 2), \\
<\sbm{p}^\prime_1 \sbm{p}^\prime_2 \mid A^{(\omega)}_{\alpha}(\sbm{k}) \mid
\sbm{p}_1 \sbm{p}_2 >&=& C(-1)^\alpha \tau_{\alpha}^1
\frac{-1}{m_N^2}v_{\omega}(\sbm{q}) \nonumber \\
&\times&\{\sbm{\sigma}^1\cdot \sbm{P}_2
+\frac{i}{2}\sbm{\sigma}^1\times\sbm{\sigma}^2\cdot\sbm{q}\}
+( 1 \leftrightarrow 2 ), \\
<\sbm{p}^\prime_1 \sbm{p}^\prime_2 \mid A^{(\rho)}_{\alpha}(\sbm{k}) \mid
\sbm{p}_1 \sbm{p}_2 >&=& C(-1)^\alpha
\frac{-1}{m_N^2}v_{\rho}(\sbm{q}) \nonumber \\
&\times&\{\tau^2_\alpha \{\sbm{\sigma}^1
\cdot \sbm{P}_2
+\frac{i}{2}\sbm{\sigma}^1\times\sbm{\sigma}^2\cdot \sbm{q}
\} \nonumber \\
&-&\frac{i}{2}[\sbm{\tau}^1\times\sbm{\tau}^2]_\alpha \sbm{\sigma}^1
\cdot \sbm{q}\} + ( 1 \leftrightarrow 2 ).
\end{eqnarray}
Here we have used the notation
\begin{eqnarray}
v_{i}(\sbm{q}) =\frac{S_i g^2_i}{\sbm{q}^2+m_i^2}F_i(\sbm{q})^2.
\end{eqnarray}
where  $S_i = -1$ for $i=\sigma,\delta$, $S_i=+1$ for $i=\omega,\rho$
, and the
$g_i$'s are the corresponding meson-nucleon-nucleon
coupling constants . In the case of the Bonn potential model [15]
the form factors and coupling constants are determined
by a fit to nucleon-nucleon scattering data (We here use the
parameters values in Table A.3 of Ref.[15]). Note that the
terms due to the tensor coupling of $\rho$ and $\omega$ vector meson
were found [3] to be unimportant and hence are neglected here.

For a general nucleon-nucleon potential, such as the Paris potential [16]
used in this work, the derivation of axial charge operator can be
done
[3,14] by the expansion of the potential in terms of Fermi invariants. The
resulting expressions are the same as above except that the radial functions
$v_{\sigma}(q), v_{\delta}(q), v_{\omega}(q)$, and $ v_{\rho}(q)$ are replaced
respectively by $v_S^{+}(q), v^-_S(q), v_V^{+}(q)
$, and $v_V^{-}(q)$. In this work we consider the Paris potential[16] and
use the $v_S^{\pm}(q)$ and $v_V^{\pm}(q)$
calculated in Ref.[14].

The above expressions are used in the numerical calculations
of $\pi$ absorption
on $^3He$ at threshold reported below.

\vspace{0.4cm}
\centerline{\bf 3. Transition rate for the reaction $^3He(\pi,NN)$ }

The kinematics of the reaction $^3He(\pi,NN)$ in the $^3He$ rest frame
is illustrated in Fig.2.
As we are interested only in the ratio between the absorption rates,
which is not expected to depend sensitively on the wavefunctions, we
shall neglect the nuclear distortion effect on the pion wavefunction.
For simplicity the $^3He$ wavefunction is taken to be
of a simple harmonic oscillator form with the oscillator parameter
value $b = 1.36$ $fm$. It then
has the separable form $ \sim \Phi (\sbm{P})\phi_{S,T}(\sbm{p})$, where
$\sbm{P}$
and $\sbm{p}$ are the total and the relative momentum of
a pair of nucleons in $^3He$ respectively. With these simplifications and
the kinematic variables shown in Fig.2,
the total transition rate of
the reaction $^3He(\pi, NN)$
is then determined by the following quantity
\begin{eqnarray}
P_{m_{\tau_1}m_{\tau_2};\alpha}
&=&\frac{3}{2j+1}\sum_{m_j}\sum_{m_{s_1},m_{s_2}}
\int d\sbm{P} d\sbm{p} \delta (\omega_{\pi}(\sbm{k})
- \epsilon - \frac{P^2}{4m_N}
-\frac{p^2}{m_N})\mid \Phi(\sbm{P})\mid ^2 \nonumber \\
&\times& \mid \sum_{S,T} \sum_{M_S,M_T}
<S s M_S m_{s_3}\mid j m_j >
< T \tau M_T m_{\tau_3}\mid t m_t > \nonumber \\
&\times& < \sbm{p}_1 m_{s_1} m_{\tau_1}, \sbm{p}_2 m_{s_2}
m_{\tau_2} \mid A \mid \phi^{M_S,M_T}_{S,T}, \sbm{k},\alpha >\mid ^2,
\nonumber \\
\end{eqnarray}
where $\tau=s=1/2$,
$j=1/2$ and $t=m_t=1/2$ are the spin and isospin quantum numbers of
$^3He$, and obviously $m_{s_3} = m_j-M_S, m_{\tau_3} = m_t-M_T$.
The isospin quantum numbers and the
momenta of the ejected two nucleons are
$(m_{\tau_1},m_{\tau_2})$ and
$\sbm{p}_1=\sbm{P}+\sbm{p}/2, \sbm{p}_2=\sbm{P}-\sbm{p}/2$.
The momentum and the charge of the pion are respectively denoted $\sbm{k}$
and $\alpha$. The outgoing two-nucleon state is properly antisymmetrized.

In the stopped pion limit $\sbm{k}\rightarrow 0$ Eq.(23)
can be written in the following
partial-wave decomposed form
\begin{eqnarray}
P_{m_{\tau_1}m_{\tau_2};\alpha} &=&\sum_{\gamma,\beta}
<T^\prime M^\prime _T\mid \tau\tau m_{\tau_1} m_{\tau_2}>
<T \tau M_T m_{\tau_3}\mid t m_t>\nonumber \\
&\times&<T^\prime M^\prime _T\mid T 1 M_T \alpha > g(\gamma,\beta),
\end{eqnarray}
where
\begin{equation}
g(\gamma,\beta)=\frac{3}{2}\int d\sbm{P}\mid\Phi (\sbm{P})\mid^2
\mid <\gamma,p\mid A\mid \phi_\beta>\mid^2,
\end{equation}
with $p=(m_\pi-\epsilon-3P^2/4M_N)^{1/2}$.
Here we have introduced abbreviated partial-wave
notations $\beta : [LS]JT$ for the initial two-nucleon bound state, and
$\gamma :[L^\prime S^\prime]J^\prime T^\prime$ for the ejected two nucleons.
The matrix element of $A$ has contributions
from one-body term and two-body terms defined in section 2
\begin{eqnarray}
<\gamma,p\mid A\mid \phi_{\beta}>&= &<\gamma,p\mid A^{(1)}\mid \phi_{\beta}>
\nonumber \\
&+& \sum_{i}<\gamma,p\mid A^{(i)}\mid \phi_{\beta} >
\end{eqnarray}
where $i=\pi,\sigma,\delta,\rho,\omega$.
The contribution from the one-body term is
\begin{eqnarray}
<\gamma,p^\prime \mid A_{\alpha}^{(1)} \mid \Phi_\beta>
=-B(2\pi)^3\frac{f_{\pi NN}}{m_\pi}\frac{p}{m}\phi_{[LS]JT}(p),
\end{eqnarray}
where $\phi_{[LS]JT}(p)$ is a general bound state wavefunction. For an $S$-wave
harmonic oscillator wavefunction of $^3He$, we have $L=0, J=S$ and
only $[S,T]=[0,1],[1,0] $ are permitted by the Pauli principle.
The radial wavefunction is $\phi_{[0,S]S,T} (p) = \phi (p)=
N e^{-b^2p^2}$. The angular momentum coupling in Eq.(27) is isolated
in the coefficient $B$. We find that
\begin{eqnarray}
B &=& i \sqrt{\frac{m_{\pi}}{2}}\frac{1}{(2\pi)^{9/2}}\hat{L}\hat{L^\prime}
\hat{S^\prime}(-1)^{(J-S^\prime+1)}\delta_{J^\prime J}\delta_{M M^\prime}
\nonumber \\
&\times& \left(
\begin{array}{ccc}
1 &L^\prime& L \\
0 &0&        0
\end{array}
\right)
\left\{
\begin{array}{ccc}
S &L& J \\
L^\prime &S^\prime& 1
\end{array}
\right\}.
\end{eqnarray}
The matrix element of the pion $S$-wave rescattering term is
\begin{eqnarray}
<\gamma,p^\prime \mid
A_{\alpha}^{(\pi )} \mid \phi_\beta > =
B\frac{8\pi f_{\pi NN}}
{m_\pi}\frac{1}{m_\pi^2}[A^{(\pi)}_{L^\prime L}(L^\prime )
-B^{(\pi)}_{L^\prime L}(L)] \nonumber \\
\times\left\{\lambda_1 <S^\prime T^\prime \mid\mid O_1 \mid \mid ST>
+\frac{\omega_{\pi}(k)+m_\pi }{m_{\pi}}\lambda_2
<S^\prime T^\prime \mid\mid O_3 \mid\mid ST > \right\},
\end{eqnarray}
where
\begin{eqnarray}
A^{(\pi)}_{L^\prime L}(l) =\int p^2 dp F^{(\pi)}_l(p^\prime,p)\phi_{[LS]JT}
(p)p,
\nonumber \\
B^{(\pi)}_{L^\prime L}(l) = p^\prime \int p^2 dp F^{(\pi)}_l(p^\prime,p)
\phi_{[LS]JT}(p),
\end{eqnarray}
with
\begin{eqnarray}
F^{(\pi)}_l(p^\prime,p)=(2\pi)\int_{-1}^{+1} P_l(x) d x
v_{\pi}(\sbm{p}^\prime-\sbm{p}),
\end{eqnarray}
where $x=\hat{p}^\prime\cdot\hat{p}$.
Defining similar partial-wave projected integrals for the
$i=\sigma,\delta,\rho$ and $\omega$ mesons exchange contributions,
we then get the following expressions
\begin{eqnarray}
<\gamma,p^\prime \mid
A_{\alpha}^{(\sigma)} \mid \phi_\beta >&=& B\frac{g_A}{2f_\pi}
\frac{1}{2m_N^2}\left\{
A^{(\sigma)}_{L^\prime L}(L^\prime) +B^{(\sigma)}_{L^\prime L}(L)\right\}
\nonumber \\
&\times& < S^\prime T^\prime \mid\mid O_1 \mid\mid ST >,
\end{eqnarray}
\begin{eqnarray}
<\gamma,p^\prime \mid
A_{\alpha}^{(\delta)} \mid \phi_\beta >&=&
B\frac{g_A}{2 f_\pi}\frac{1}{2m_N^2}
\left\{A^{(\delta)}_{L^\prime L}(L^\prime)+B^{(\delta)}_{L^\prime L}(L)\right\}
\nonumber \\
&\times& <S^\prime T^\prime \mid\mid O_2 \mid\mid ST >,
\end{eqnarray}
\begin{eqnarray}
<\gamma, p^\prime \mid
A_{\alpha}^{(\omega)} \mid \phi_\beta >= B\frac{g_A}{2f_\pi}
\frac{1}{2m_N^2} \nonumber \\
\times
\{
 -(A^{(\omega)}_{L^\prime L}(L^\prime)+B^{(\omega)}_{L^\prime L}(L))
<S^\prime T^\prime \mid\mid O_1\mid\mid ST> \nonumber \\
+\sqrt{2}(A^{(\omega)}_{L^\prime L}(L^\prime)-B^{(\omega)}_{L^\prime L}(L))
<S^\prime T^\prime \mid\mid O_4 \mid\mid ST > \},
\end{eqnarray}
\begin{eqnarray}
<\gamma,p^\prime \mid
A_{\alpha}^{(\rho)} \mid \phi_\beta >&=& B\frac{g_A}{2f_\pi}
\frac{(1+\kappa)}{2m_N^2} \nonumber \\
&\times&\{-(A^{(\rho)}_{L^\prime L}(L^\prime)
+B^{(\rho)}_{L^\prime L}(L))<S^\prime T^\prime \mid\mid O_2 \mid\mid ST>
\nonumber \\
&+&\sqrt{2}(A^{(\rho)}_{L^\prime L}(L^\prime)-B^{(\rho)}_{L^\prime L}(L))
(<S^\prime T^\prime \mid\mid O_4 \mid\mid ST> \nonumber \\
&+&<S^\prime T^\prime \mid\mid O_3 \mid\mid ST>) \} .
\end{eqnarray}
The reduced matrix elements in the above equations
are defined by the following spin-isospin
operators
\begin{eqnarray}
\sbm{O}_1 = \sbm{\tau}^1\sbm{\sigma}^1 -\sbm{\tau}^2\sbm{\sigma}^2,
\nonumber \\
\sbm{O}_2 = \sbm{\tau}^2\sbm{\sigma}^1-\sbm{\tau}^1\sbm{\sigma}^2,
\nonumber \\
\sbm{O}_3 = \frac{i}{\sqrt{2}}\sbm{\tau}^1\times \sbm{\tau}^2
(\sbm{\sigma}^1+\sbm{\sigma}^2) \nonumber \\
\sbm{O}_4 = (\sbm{\tau}^1+\sbm{\tau}^2)\frac{i}{\sqrt{2}}
\sbm{\sigma}^1\times\sbm{\sigma}^2.
\end{eqnarray}
It is then easy to show that
\begin{eqnarray}
<S^\prime T^\prime \mid\mid \sbm{O}_1 \mid\mid ST > &=& 6 \hat{S}\hat{T}
((-1)^{(S+T)} -(-1)^{(S^\prime +T^\prime)}) \nonumber \\
&\times&\left\{
\begin{array}{ccc}
T^\prime &  T  & 1 \\
1/2      & 1/2 & 1/2
\end{array}\right\}
\left\{
\begin{array}{ccc}
S^\prime &S& 1 \\
1/2 &1/2& 1/2
\end{array}\right\},
\end{eqnarray}
\begin{eqnarray}
<S^\prime T^\prime \mid\mid \sbm{O}_2 \mid\mid ST> &=& 6\hat{S}\hat{T}
((-1)^{(S+T^\prime)}-(-1)^{(S^\prime +T)}) \nonumber \\
&\times&
\left\{
\begin{array}{ccc}
T^\prime &T& 1 \\
1/2 &1/2& 1/2
\end{array}\right\}
\left\{
\begin{array}{ccc}
S^\prime &S& 1 \\
1/2 &1/2& 1/2
\end{array}\right\},
\end{eqnarray}
\begin{eqnarray}
<S^\prime T^\prime \mid\mid \sbm{O}_3 \mid\mid ST >& =&6
\sqrt{18}\hat{S}\hat{T}
((-1)^S+(-1)^{S^\prime}) \nonumber \\
&\times&\left\{
\begin{array}{ccc}
T^\prime &T& 1 \\
1/2 &1/2& 1 \\
1/2 &1/2& 1
\end{array}\right\}
\left\{
\begin{array}{ccc}
S^\prime &S& 1 \\
1/2 &1/2& 1/2
\end{array}\right\},
\end{eqnarray}
\begin{eqnarray}
<S^\prime T^\prime \mid\mid \sbm{O}_4 \mid\mid ST> &=& 6\sqrt{18}\hat{S}\hat{T}
((-1)^T+(-1)^{T^\prime}) \nonumber \\
&\times&\left\{
\begin{array}{ccc}
S^\prime &S& 1 \\
1/2 &1/2& 1 \\
1/2 &1/2& 1
\end{array}\right\}
\left\{
\begin{array}{ccc}
T^\prime &T& 1 \\
1/2 &1/2& 1/2
\end{array}\right\} .
\end{eqnarray}
\vspace{0.4 cm}

\centerline{\bf 4. Results and Discussion}

It is interesting to first note that all transition matrix elements
, Eq.(27) and Eqs.(29)-(35), depend on the same
geometric factor $B$ defined in Eq.(28). Evaluating this factor for
an initial $S$-wave ($L =0$) NN pair,
it is easy to see that only the following transitions can occur in the
reaction $^3He(\pi,NN)$:
\begin{eqnarray}
f_{11}&=&g(^3P_0,^1S_0) \nonumber \\
f_{10}&=&g(^3P_1,^3S_1). \nonumber
\end{eqnarray}
Here the rate factor $g$ is that defined in Eq.(25), and
we use the standard spectroscopic notation $^{2S+1}L_J$. The
subindices on the $f^\prime$s indicate the total isospin $T^\prime$ and
$T$ of the final and initial two-nucleon systems. Evaluation of the
Clebsch-Gordon coefficients in Eq.(24) leads to the following
ratio of the absorption probabilities
($P_{m_{\tau_1}m_{\tau_2};\alpha}$ defined in Eq.(24))
for negative pions :
\begin{eqnarray}
R&=&\frac{^3He(\pi^-,nn)}{^3He(\pi^-,np)}=
\frac{P_{-1/2,-1/2;-1}}{P_{-1/2,1/2;-1}} \nonumber \\
&=& 1 + 6\frac{f_{10}}{f_{11}} .
\end{eqnarray}

The pion rescattering term $A^{(\pi)}$, Eq.(29), gives a
very small contribution to
$f_{11}$ because its main ($\lambda_2$) term depends on
the isospin operator $\sbm{\tau}^1\times\sbm{\tau}^2$
and therefore its matrix element between two
$T=1$ states vanishes. The remaining term gives only a small contribution
because the coefficient $\lambda_1 \simeq 0$, as discussed in section 3.
On the other hand, the pion rescattering term gives a large contribution to
the transition between $T=1$ and $T=0$ states ( i.e. $f_{10}$) and hence
the prediction based on the sum of
the single nucleon contribution $A^{(1)}$ and the pion rescattering term
$A^{(\pi)}$ is very large, as shown in the first row of Table 1.

In Table 1 we also show the effects caused by the short-range absorption
mechanisms as calculated using the Bonn [15] and Paris [16] potential
models. By comparing rows 1-3, it is seen that the short
range corrections due to
the $\sigma$ and $\omega$ exchange mechanisms (or more generally from the
isospin independent scalar and vector exchange mechanisms) drastically
reduce the predicted ratio.
The contributions from isovector $\delta$ and $\rho$ exchanges are less
important, but are significant in further reducing the predicted ratio
towards the experimental value [7].
Although the contributions from the individual two-nucleon mechanisms
differ between the two potential models,  the net result obtained by
taking all of them into account are quite similar and close to
the experimental value. This situation is in agreement with that found for
the reaction
$pp \rightarrow pp \pi^0$ near threshold [1].

The results presented in Table 1 further strengthens the case
for the importance of the short range
pion absorption mechanisms [1].
It will be interesting to explore whether this model also is able
to resolve several long-standing problems associated with nuclear
pion absorption at higher energies. In the case of the two-nucleon system,
all existing unitary $\pi NN$ calculations [17] based on pion rescattering
mechanisms ( including $S$ and $P$ $\pi N$ partial waves) fail to give
satisfactory descriptions of the analyzing powers for the
reactions $\pi^+ d \rightarrow pp$ and $NN \rightarrow \pi NN$.
Although the short-range absorption mechanisms that involve intermediate
$N\bar N$pairs ( Fig.1) is
in general weaker than the pion rescattering term, it can
nevertheless have
large effects on the polarization observables through interference
with the large pion rescattering amplitude. This can be explored
straightforwardly within the unitary
$\pi NN$ formalism developed by Lee and
Matsuyama [18].

It will be also interesting to investigate the effect of
the short-range absorption mechanisms on $^3He(\pi^{-}, np)$ reaction
in the $\Delta$ region. This reaction involves small contribution from
the $\Delta$ excitation, since the dominant $N \Delta$ $S$-wave intermediate
state is excluded [17] for an initial $^1S_0$ pp pair in $^3He$. The
angular distributions of this reaction as well as the energy-dependence
of the ratio of $^3He(\pi^+, pp)/^3He(\pi^-,np)$ have not been well
understood [17].
An extension of the theoretical calculations, such that developed in
Ref.[18], to include the short-range absorption mechanisms
is expected to significantly change the theoretical predictions of these
absorption observables for $^3He$.
A rigorous theoretical calculation of
two-body absorption on $^3He$ is essential for extracting from the
data the contribution from three-body absorption mechanisms which have
been empirically established in Ref.[17].

\vspace{2.0 cm}
\centerline{\bf Acknowledgements}

We thank H. Weyer and his collaborators for informing us of their data
on the absorption of stopped $\pi^-$ on $^3He$.
This work was supported in part by the U.S. Department of Energy,
Nuclear Physics Division, under Contract No. W-31-109-ENG-38,
and in part by the exchange program between the U.S.
National Science Foundation
and the Academy of Finland. The support to L.L.K. by
Argonne Theory Visitor Program funded by Argonne Associate
Laboratory Director for Physical Research is acknowledged.

\newpage
{\bf References}
\begin{enumerate}

\item T.-S. H. Lee and D. O. Riska, Phys. Rev. Lett. {\bf 70}, 2237 (1993)
\item K. Kubodera, J. Delorme and M. Rho, Phys. Rev. Lett. {\bf 40}, 755 (1978)
\item M. Kirchbach, D. O. Riska and K. Tsushima, Nucl. Phys. {\bf A542}, 616
(1992)
\item H. O. Meyer et al , Phys. Rev. Lett. {\bf 65}, 2846 (1990)
      Nucl. Phys. {\bf A539}, 633 (1992)
\item C. J. Horowitz, H. O. Meyer and D. K. Griegel,
Phys. Rev. {\bf C49}, 1337(1994)
\item J. A. Niskanen, Phys. Rev. {\bf C49}, 1285(1994)
\item D. Gotta et al., Preprint Univ. Basel 1994
\item S. Weinberg, Phys. Rev. Lett. {\bf 18}, 188 (1967)
\item E. M. Nyman and D. O. Riska, Phys. Lett. {\bf 215}, 29 (1988)
\item A. E. Woodruff, Phys. Rev. {bf 117}, 1113 (1960)
\item D. Koltun and A. Reitan, Nucl. Phys. {\bf B4}, 629 (1968)
\item G. H\"{o}hler et al., Handbook of Pion-Nucleon Scattering,
Physics Data 12-1, Karlsruhe (1979)
\item R. A. Arndt, J. M. Ford and L. D. Roper, Phys. Rev. {\bf D32}, 1085
(1985)
\item P. G. Blunden and D. O. Riska, Nucl. Phys. {\bf A536}, 697 (1992)
\item R. Machleidt, Advances in Nuclear Physics, J.W. Negele and E. Vogt
eds. Plenum, New York (1989)
\item M. Lacombe et al., Phys. Rev. {\bf C21}, 861 (1980)
\item H. Garcilazo and T. Mizutani, $\pi NN$ Systems, World Scientific pub.
      Co.(1990)
\item A. Matsuyama and T.-S. H. Lee, Phys. Rev. {\bf C34}, 1900 (1986);
      Nucl. Phys. {\bf A526}, 547(1991)
\item D. Ashery and J. P. Schiffer, Ann. Rev. Nucl. Part. Sci., {\bf}36,
207 (1986); H. Weyer, Phys. Reports {\bf 195}, 195 (1990)
\item K. Ohta, M. Thies and T.-S. H. Lee, Ann. Phys. {\bf 163}, 420 (1985)

\end{enumerate}

\newpage
\centerline{\bf Table 1}

\vspace*{1cm}

The ratios $R = \,^3He(\pi^-,nn)/^3He(\pi^-,np)$ as predicted using the
Bonn [11] and Paris [12] potential models in the construction of the
two-nucleon operators are compared with the experimental value [7].
The absorption mechanisms included in each calculation,
defined in Eq.(27) and Eqs.(29)-(35)
, are indicated in each row.
\vspace*{1cm}

\begin{tabular}{|c|c|c|c|} \hline
Mechanism & $R_{Bonn}$ & $R_{Paris}$ & $R_{Exp.}$ \\ \hline
$A^{(1)} +A^{(\pi)}$ & 533.42 & 533.42 & -  \\ \hline
$A^{(1)} +A^{(\pi+\sigma)}$ & 74.30 & 208.12 & - \\ \hline
$A^{(1)} +A^{(\pi+\sigma+\omega)}$ & 12.40 & 16.81 & - \\ \hline
$A^{(1)} +A^{(\pi+\sigma+\omega+\delta)}$ & 11.77 & 11.88 & -  \\ \hline
$A^{(1)} +A^{(+\pi +\sigma + \omega + \delta+ \rho}) $ & 9.786
& 8.0349 & 6.3 $\pm$  1.1
\\ \hline
\end{tabular}
\vspace{2cm}

\centerline {\bf Figure Captions}

{\bf Fig.~1: } Axial exchange current operator arising from
intermediate $N\bar N$ pairs that contributes to the
$ NN \leftrightarrow NN \pi$
reaction.
\noindent

{\bf Fig.~2: } Kinematics of $^3He(\pi, NN)$ reaction in the rest frame of
$^3He$.

\newpage
\end{document}